\documentstyle[12pt]{article}
\begin{document}

\author{{\bf R. Bertolotti, G. Contreras, L. A. N\'{u}\~{n}ez}\thanks{%
nunez@ula.ve}, {\bf U. Percoco}\thanks{%
e-mail: upercoco@dino.conicit.ve}\thanks{%
On leave from Centro de Fisica, Instituto Venezolano de Investigaciones
Cientificas (IVIC) Apartado 21827, Caracas 1020-A, Venezuela.} \\
%EndAName
{\it Departamento de F\'\i sica, Facultad de Ciencias, }\\
{\it \ Universidad de los Andes, 5101 M\'erida, Venezuela.} \and {\bf J.
Carot}\thanks{%
e-mail: dfsjcg0@ps.uib.es} \\
%EndAName
{\it Departmento de F\'\i sica, Universitat de les Illes Baleares }\\
{\it \ E-07071, Palma de Mallorca, Espa\~na.}}
\title{{\bf Comment on Ricci Collineations of Static Spherically Symmetric
Spacetimes.}}
\maketitle

\begin{abstract}
We present a counter example to a theorem given by Amir {\em et al.} J.
Math. Phys. {\bf 35}, 3005 (1994). We also comment on a misleading statement
of the same reference.
\end{abstract}

In a recent paper, M. Jamil Amir {\em et al.} \cite{pak} have presented a
detailed analysis of Ricci collineations (RC) for static, spherically
symmetric spacetimes, with a special focus on the relationship between RC
and isometries (KV). This has led the authors to the following theorem:\\%
{\bf Theorem 1}: {\it Spherically symmetric static spacetimes with nonzero
(and non infinite) determinant of the Ricci tensor have RCs identical with
the Killing Vector, but when the determinant is zero there may be additional
degrees of freedom, giving infinitely many RCs for each degree of freedom.}

We have found a counter example to this result. The metric for spherically
symmetric static spacetimes can be written in the form\cite{kramer}
\begin{equation}
ds^2=-\ e^{\nu (r)}\,dt^2+\,e^{\lambda (r)}\,dr^2+\,r^2(d\vartheta ^2+\,\sin
^2\vartheta \,d\phi ^2)  \label{1}
\end{equation}

The Ricci tensor for this metric is diagonal and can be written as: $%
R_{00}~=~A(r)$, $R_{11}~=~B(r),\ R_{22}=C(r)$ and$\ R_{33}=C(r)\sin
^2\vartheta $. Let us consider the particular case in which $A(r)=$ $C(r)=1$%
, which leads to the metric

\begin{eqnarray}
\nu (r) &=&\frac{r^4}{8r_0^2}+h\,\ln \frac r{r_0}+k  \label{metric} \\
\lambda (r) &=&\nu (r)+2\,\ln \frac r{r_0}
\end{eqnarray}

with

\begin{equation}  \label{b}
B(r)=2\frac{h+1}{r^2}
\end{equation}
where $r_0$, $h$ and $k$ are constants.

{}From the equation for RC
\begin{equation}  \label{2}
\pounds _\xi \,R_{ab}=0
\end{equation}
we obtain,

\begin{equation}
\xi _{,t}^t=0  \label{LieRicci00}
\end{equation}
\begin{equation}
\xi _{,r}^t+B(r)\xi _{,t}^r=0  \label{LieRicci01}
\end{equation}
\begin{equation}
\xi _{,\vartheta }^t+\xi _{,t}^\vartheta =0  \label{LieRicci02}
\end{equation}
\begin{equation}
\xi _{,\phi }^t+\sin ^2\vartheta \ \xi _{,t}^\phi =0  \label{LieRicci03}
\end{equation}
\begin{equation}
B^{\prime }(r)\xi ^r+2B(r)\ \xi _{,r}^r=0  \label{LieRicci11}
\end{equation}
\begin{equation}
B(r)\xi _{,\vartheta }^r+\xi _{,r}^\vartheta =0  \label{LieRicci12}
\end{equation}
\begin{equation}
B(r)\xi _{,\phi }^r+\sin ^2\vartheta \xi _{,r}^\phi =0  \label{LieRicci13}
\end{equation}
\begin{equation}
\xi _{,\vartheta }^\vartheta =0  \label{LieRicci22}
\end{equation}
\begin{equation}
\xi _{,\phi }^\vartheta +\sin ^2\vartheta \ \xi _{,\vartheta }^\phi =0
\label{LieRicci23}
\end{equation}
\begin{equation}
\cot \vartheta \ \xi ^\vartheta +\xi _{,\phi }^\phi =0  \label{LieRicci33}
\end{equation}
Equations (\ref{LieRicci00}) and (\ref{LieRicci11}) can be integrated,
giving $\xi ^t~=~\Sigma (r,\vartheta ,\phi )$ and

$\xi ^r~=~K(t,\vartheta ,\phi )\,B^{-1/2}$ respectively. Substituting these
expressions into the $\vartheta $- derivative of eq. (\ref{LieRicci01}) we
find that $K(t,\vartheta ,\phi )\,=S_1(\vartheta ,\phi )~t+S_2(\vartheta
,\phi )$. Using these results into (\ref{LieRicci02}) and (\ref{LieRicci03}%
), we obtain an expression for $\xi $:

\begin{equation}  \label{psitr-1}
\xi ^t=\Sigma (r,\vartheta ,\phi ),\quad \xi ^r=\frac{S_1(\vartheta ,\phi
)\,\,t+S_2(\vartheta ,\phi )}{B^{1/2}}
\end{equation}

\begin{equation}  \label{psith-1}
\xi ^\vartheta =-\Sigma _{,\vartheta }\,\,t+\Gamma (r,\vartheta ,\phi
),\quad and\quad \xi ^\phi =-\Sigma _{,\phi }\,\,t+\Psi (r,\vartheta ,\phi )
\end{equation}

Substitution of $\xi $ into (\ref{LieRicci01}) and (\ref{LieRicci12})-(\ref
{LieRicci33}) enables us to completely determine the functions $\Sigma
(r,\vartheta ,\phi )$, $S_1(\vartheta ,\phi )\,$, $S_2(\vartheta ,\phi )\,$,
$\Gamma (r,\vartheta ,\phi )$ and $\Psi (r,\vartheta ,\phi )$. Then,

\begin{eqnarray}
\xi ^0 &=&-\,c_4\sqrt{2\,\left( h+1\right) }\,\ln r+c_0  \nonumber \\
\xi ^1 &=&\frac{c_4\,t+c_5}{\sqrt{2\,\left( h+1\right) }}r  \label{psi} \\
\xi ^2 &=&c_1\sin \phi -c_2\cos \phi  \nonumber \\
\xi ^3 &=&\left( c_1\cos \phi +c_2\sin \phi \right) \cot \vartheta +c_3
\nonumber
\end{eqnarray}

According to Theorem 1, (\ref{psi}) should represent an isometry; however it
is easy to see that $\xi $ does not reduce to a KV unless $c_4=c_5=0$. This
result invalidates the theorem stated above. Moreover, it is
straightforward, but tedious, to show that the same condition is necessary
for $\xi $ to reduce to a Riemann collineation $($\pounds $_\xi
\,R_{bcd}^a=0)$. Therefore (\ref{psi}) is a proper RC.

\qquad Next, we should like to make a remark on a misleading statement that
appears in Amir {\em et al}. \cite{pak}: ``... there is no reason, {\em a
priori}, why a RC should be a KV or {\em vice versa.}..'' . It is easy to
show that any KV\ is a RC. Indeed, Katzin {\em et al.}\cite{katzin} have
proved that a necessary an sufficient condition for a spacetime to admit a
curvature collineation (CC)\ is

\begin{equation}  \label{20}
\begin{array}{c}
\left( (\pounds _\xi \,g_{im})_{;j}+(\pounds _\xi \,g_{mj})_{;i}-(\pounds
_\xi \,g_{ij})_{;m}\right) _{;k}- \\
\qquad \qquad \left( (\pounds _\xi \,g_{km})_{;j}+(\pounds _\xi
\,g_{mj})_{;k}-(\pounds _\xi \,g_{kj\,})_{;m}\right) _{;i}=0
\end{array}
\end{equation}

Then, it is evident that a KV satisfies the above condition and, as it is
well known, every CC is a RC.

The analysis of RC for the general case of static, spherically symmetric
metrics, i.e. $A(r)$ and $C(r)$ arbitrary, is obviously more involved and
its complete solution will be reported elsewhere. The study of RC for {\em %
nonstatic} spherically symmetric metrics gives interesting results which
will be the subject of a subsequent publication.

This work has been partially supported by the Consejo de Desarrollo
Cient\'\i fico Human\'\i stico y Tecnol\'ogico de la Universidad de Los
Andes; and the Programa de Formaci\'on e Intercambio Cient\'\i fico (Plan
III). Algebraic calculations of the present work has been checked with MAPLE
V. The authors wish to thank the staff of the {\em SUMA}, the computational
facility of the Faculty of Science (Universidad de Los Andes), for making
this work possible.

\end{document}